\documentstyle[epsfig]{mn}
\def\etal{{\it et al.}\thinspace}

\def\bi{\bibitem{}}

\def\ni{\noindent}
\def\beb{\begin{thebibliography}{999}}
\def\eeb{\end{thebibliography}}
\def\bei{\begin{itemize}}
\def\eei{\end{itemize}}
\def\bef{\begin{figure}}
\def\eef{\end{figure}}
\def\ben{\begin{enumerate}}
\def\een{\end{enumerate}}
\def\beq{\begin{equation}}
\def\eeq{\end{equation}}
\def\ber{\begin{eqnarray}}
\def\eer{\end{eqnarray}}

\newcommand{\gcc}{{\rm g} \, {\rm cm}^{-3}}

\newcommand{\msun}{\mbox{{\rm M}$_{\odot}$}}

\newcommand{\lsim}{\raisebox{-0.3ex}{\mbox{$\stackrel{<}{_\sim} \,$}}}
\newcommand{\gsim}{\raisebox{-0.3ex}{\mbox{$\stackrel{>}{_\sim} \,$}}}
\begin{document}
\title[neutron star field evolution]
{Evolution of Multipolar Magnetic Field in Isolated Neutron Stars}
\author[Mitra, Konar and Bhattacharya]
{Dipanjan Mitra$^{1,{\ast}}$, Sushan Konar$^{2,{\ast}}$ and Dipankar Bhattacharya$^{1,{\ast}}$ \\
$^1$Raman Research Institute , Bangalore 560080, India \\
$^2$Inter-University Centre for Astronomy and Astrophysics, Pune 411007, India \\ 
$^{\ast}$ e-mail : dmitra@rri.ernet.in, sushan@iucaa.ernet.in, dipankar@rri.ernet.in}
\date{9th October, 1998}
\maketitle
\baselineskip=10pt

\begin{abstract}
The evolution of the multipolar structure of the magnetic field of isolated 
neutron stars is studied assuming the 
currents to be confined to the crust. We find that except for multipoles 
of very high order ($l\gsim 25$) the evolution 
is similar to that of a dipole.  Therefore no significant evolution is expected in pulse
shape of isolated radio pulsars due to the evolution of the multipole structure of 
the magnetic field.
\end{abstract}

\begin{keywords}
magnetic fields: multipole--stars: neutron--pulsars: general
\end{keywords}

\section{Introduction}

\ni Strong multipole components of the magnetic field have long been 
thought to play an important role in the radio 
emission from pulsars. Multipole fields have been 
invoked for the generation of electron positron pairs in the pulsar 
magnetosphere. For example, the Ruderman \& Sutherland (1975) model requires 
that the radius of curvature of the field lines 
near the stellar surface should be of the order of the stellar 
radius to sustain pair production in long period pulsars. 
This is much smaller than the expected radius of curvature of the dipole 
field. Such a small radius of curvature could be a signature of either
an extremely offset dipole (Arons, 1998) or of a multipolar 
field (Barnard \& Arons, 1982).
Further, soft X-ray observations
of pulsars show non-uniform surface temperatures which can be attributed
to the presence of a quadrupolar field (Page \& Sarmiento, 1996).  \\

\ni Magnetic multipole structure at and near the polar cap is also 
thought to be responsible for the unique 
pulse profile of a pulsar (Vivekanand \& Radhakrishnan 
1980, Krolik 1991, Rankin \& Rathnasree 1995). The 
recent estimates that there should be several tens of sparks populating 
the polar cap is also best explainable 
if multipole fields dictate the spark geometry near the 
surface (Deshpande \& Rankin 1998, Rankin \& Deshpande 1998, Seiradakis 1998). 
Significant evolution in the 
structure of the magnetic field during the lifetime of a pulsar may 
therefore leave observable signatures. If 
the multipoles grow progressively weaker in comparison to 
the dipole then one can expect pulse profiles to simplify 
with age and vice versa. \\

\ni The evolution of the magnetic fields in neutron stars 
in general is still a relatively open question. 
During the last decade, two major alternative scenarios 
for the field evolution have emerged. One of these 
assumes that the field of the neutron star permeates the whole 
star at birth, and its evolution is dictated 
by the interaction between superfluid vortices (carrying angular momentum) 
and superconducting fluxoids 
(carrying magnetic flux) in the stellar interior. As the 
star spins down, the outgoing vortices may drag 
and expel the field from the interior leaving it to decay 
in the crust (Srinivasan \etal 1990). 
In a related model, plate tectonic motions driven by pulsar spindown 
drags the magnetic poles together, reducing
the magnetic moment (Ruderman 1991a,b,c).\\

\ni The other scenario assumes that most of the field is generated 
in the outer crust after the birth of the 
neutron star (Blandford, Applegate \& Hernquist 1983). The later evolution 
of this field is governed entirely 
by the ohmic decay of currents in the crustal layers. The evolution of the 
dipole field carried by such currents 
has been investigated in some detail in the recent 
literature (Sang \& Chanmugam 1987, Geppert \& Urpin 1994, Urpin \& Geppert 1995, 1996, 
Konar \& Bhattacharya 1997, 1998). These studies include field evolution 
in isolated neutron stars as well as 
those accreting from their binary companions. The results show 
interesting agreements with observations lending 
some credence to the crustal picture. \\

\ni In this paper, we explore the ohmic evolution of higher order 
multipoles in isolated neutron stars assuming the 
currents to be originally confined in the crustal region. 
Our goal is to find whether there would be any observable 
effect on the pulse shape of radio emission from isolated 
pulsars as a result of this 
evolution. In section 2 we discuss the details of the computation 
and in section 3 we present our results and
discuss the implications.

\section{computations}
The evolution of the magnetic field, due to ohmic diffusion, is 
governed by the equation (Jackson 1975) :
\beq
\frac{\partial {\bf B}}{\partial t} = - \frac{c^2}{4 \pi} {\bf \nabla} \times (\frac{1}{\sigma} 
                                        \times {\bf \nabla} \times {\bf B}),   
\label{e_induction}
\eeq
where $\sigma(r,t)$ is the electrical conductivity of the medium. Following Wendell, Van Horn \& Sargent (1987) 
we introduce a vector potential ${\bf A} = (0, 0, A_{\phi})$ assuming the field to be purely poloidal, such that:
\[S(r,\theta,t) = - r \, sin{\theta} \, A_{\phi}(r,\theta,t),\]
where $S(r, \theta, t)$ is the Stokes' stream function. $S$ can be separated in $r$ and $\theta$ in the form :
\[S(r,\theta,t) = \sum_{l\geq 1} R_{l}(r,t) \, sin{\theta} \, P_{l}^{1}(cos{\theta}),\]
where $P_{l}^{1}(cos(\theta))$ is the associated Legendre polynomial of degree one and $R_{l}$ is the multipole 
radial function. From equation (\ref{e_induction}) we obtain :
\beq
\frac{\partial^{2}R_{l}}{\partial x^{2}} - \frac{l(l+1)}{x^{2}} R_{l} = 
  \frac{4\pi R_{*}^{2}\sigma}{c^{2}} \frac{\partial R_{l}}{\partial t}
\label{e_radial}
\eeq
where $x \equiv r/R_{\ast}$ is the fractional radius in terms of the stellar radius $R_{\ast}$. The solution of this
equation with the boundary conditions :
\ber
\frac{\partial R_{l}}{\partial x} + \frac{l}{x} R_{l}&=&0, \; \; \mbox{as $x \rightarrow 1$} \nonumber \\
R_{l} &=& 0, \; \; \mbox{at $x = x_c$}
\label{e_bc}
\eer
for a particular value of $l$ gives the time-evolution of the multipole of order $l$. Here, the first condition 
matches the correct multipole field in vacuum at the stellar surface and the second condition makes the field vanish at 
the core-crust boundary (where $r = r_c$, the radius of the core) to keep the field confined to the crust.
We assume that the field does not penetrate the core in the course of evolution, as the core is likely to be superconducting. 

\subsection{Crustal Physics}

\ni The rate of ohmic diffusion is determined mainly by the electrical conductivity of the crust. The conductivity of 
the solid crust is given by \[\frac{1}{\sigma} = \frac{1}{\sigma_{\rm ph}} + \frac{1}{\sigma_{\rm imp}}\] where 
$\sigma_{\rm ph}$ is the phonon scattering conductivity, which we obtain from Itoh \etal (1984) as a function of density and 
temperature, and the impurity scattering conductivity $\sigma_{\rm imp}$ is obtained from the expressions 
given by Yakovlev \& Urpin (1980). \\

\ni We construct the density profile of the neutron star in question using the equation of state of Wiringa, Fiks \& Fabrocini
(1988) matched to Negele \& Vautherin (1973) and Baym, Pethick \& Sutherland (1971) for an assumed mass of 1.4~\msun.
As conductivity is a steeply increasing function of density and since the density in the crust spans eight orders 
of magnitude the conductivity changes sharply as a function of depth from the neutron star surface. Thus the deeper the 
location of the current distribution, the slower is the decay. \\

\ni Another important factor in determining the conductivity is the 
temperature of the crust. In absence of impurities 
the scattering of crustal electrons come entirely from the phonons 
in the lattice (Yakovlev \& Urpin 1980) and the 
number density of phonons increases steeply with temperature. 
The thermal evolution of the crust therefore plays an 
important role in the evolution of the magnetic field. 

The thermal evolution of a neutron star has been computed by 
many authors, and it is clearly seen that the inner crust 
($\rho > 10^{10} \rm{g cm^{-3}}$) quickly attains an isothermal
configuration after birth. At outer regions of the crust, the 
temperature follows an approximate relation,
\beq
T(\rho) = \left(\frac{\rho}{\rho_{b}}\right)^{1/4} T_{i},~~ \rho~\lsim~\rho_{b}
\label{temp}
\eeq
where $T_{i}$ is the temperature of the isothermal inner 
crust and $\rho_{b}$ is the density above which the crust is 
practically isothermal. As the star cools, larger fraction of 
the crust starts becoming isothermal, with $\rho_{b}$ 
being approximately given by,
\beq
\rho_{b} = 10^{10} \left(\frac{T_{i}}{10^{9}}\right)^{1.8}
\label{rho}
\eeq
The relations \ref{temp} and \ref{rho} above have been obtained 
by fitting to the radial temperature profiles published by
Gudmundsson, Pethick \& Epstein (1983). For the time evolution
of $T_i$ we use the results of Urpin \& van Riper (1993)
for the case of standard cooling (the crustal temperature $T_m$ in 
their notation corresponds to $T_i$ above). \\

\ni A third parameter that should be considered in determining conductivity is the impurity concentration. The effect 
of impurities on the conductivity is usually parametrised by a quantity $Q$, defined as 
$Q = \frac{1}{n} \sum_{i}{{n_{i}}(Z - Z_{i})^2}$, where $n$ is the total ion density, $n_i$ is the density of impurity 
species $i$ with charge $Z_i$, and $Z$ is the ionic charge in the pure lattice (Yakovlev \& Urpin 1980).
In the literature $Q$ is assumed to lie in the range 0.0 - 0.1. But statistical analyses indicate that the magnetic field 
of isolated pulsars do not undergo significant decay during the radio pulsar life time (Bhattacharya \etal 1992, Hartman 
\etal 1997, Mukherjee \& Kembhavi 1997). It has been shown (Konar 1997) that to be consistent with this impurity values 
in excess of 0.01 are not allowed in the crustal model. 

\bef
\begin{center}{\mbox{\epsfig{file=figure1a.ps,width=225pt}}}\end{center}
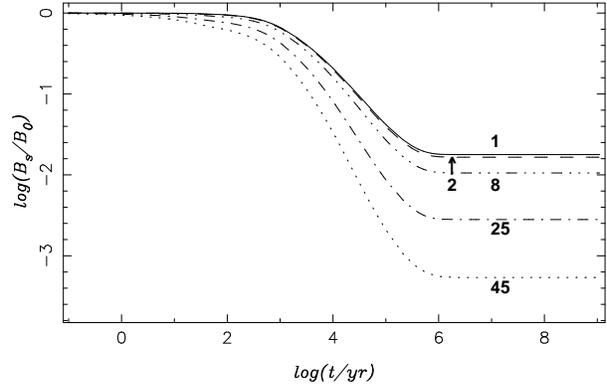
\caption[]{The evolution of the surface magnetic field for various multipoles due to pure diffusion. The numbers next to the
curves correspond to respective orders of multipole. All the curves correspond to $Q=0.0$ and a depth of current concentration
at $x_{\circ}=0.98$ i.e., a density of $\rho = 10^{11}~\gcc$.}
\label{f_fig1a}
\eef

\subsection{Numerical Scheme}
To solve equation (\ref{e_radial}) we assume the multipole radial profile used by Bhattacharya \& Datta (1996, see also
Konar \& Bhattacharya 1997). This profile contains the depth and the width of the current configuration as input parameters
and we vary them to check the sensitivity of the result to these. We solve equation (\ref{e_radial}) numerically using the 
Crank-Nicholson method of differencing. We have modified the numerical code developed by Konar (1997) and used by 
Konar \& Bhattacharya (1997) to compute the evolution of multipolar magnetic fields satisfying the appropriate boundary 
conditions given by equation (\ref{e_bc}).

\bef
\begin{center}{\mbox{\epsfig{file=figure1b.ps,width=225pt}}}\end{center}
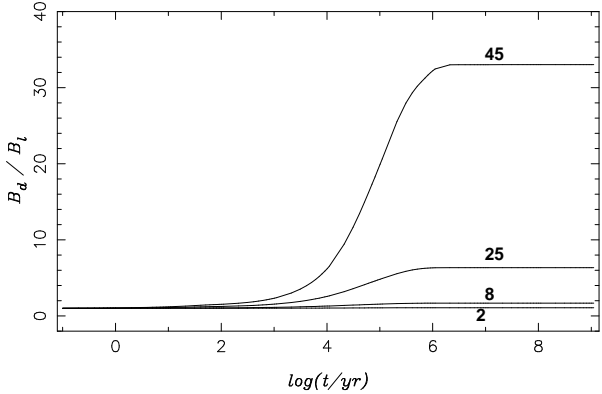
\caption[]{The ratio of the dipole surface field to the multipole field is plotted as a function of age.  The numbers next to the
curves correspond to respective orders of multipole. All the curves correspond to $Q=0.0$ and a depth of current concentration
at $x = 0.98$ i.e., a density of $\rho = 10^{11}~\gcc$.}
\label{f_fig1b}
\eef
\section{Results and Discussion}

\ni In figures [\ref{f_fig1a}] and [\ref{f_fig1b}] we plot the evolution of the various multipole components of the magnetic
field, assuming the same initial strength for all, with time due to pure diffusion in an isolated neutron star. It is evident 
from the figures that except for very high multipole orders ($l\gsim 25$) the reduction in the field strength is very similar 
to that of the dipole component. For a multipole of order $l$ there would be $2^{l}$ reversals across the stellar surface. 
For typical spin-periods the size of the polar cap bounded by the base of the open field lines is $\sim 0.01\%$ of the 
total surface area. To contribute 
to the substructure of the pulse therefore the required multipoles must have a few reversals in the polar cap which demands that 
the multipole order must be five or more. On the other hand if the multipole order is very large ($l > l_{\rm max}\sim 20$) the fine structure 
would be so small that it would be lost in the finite time resolution of observations. Therefore, $l$ values in the range 5 to $l_{\rm max}$
would be the major contributors to the observed structure of the pulse profile. However, as seen from figures [\ref{f_fig1a}] 
and [\ref{f_fig1b}] multipoles of such orders evolve similarly to the dipole. Therefore no significant evolution is expected in 
the pulse shape due to the evolution of the multipole structure of the magnetic field. As discussed before multipole orders 
contributing to the required field line curvature for pair-production are low, most prominently a quadrupole. As the evolution 
of these low orders are also very close to the dipole the radii of curvature of the field lines on the polar cap are not expected 
to change significantly in the lifetime of a radio pulsar. \\

\ni To test the sensitivity of these results on the impurity concentration of the crust and the density at which the initial
current is concentrated we have evolved models with various values of these parameters. The results are displayed in
figures [\ref{f_fig2a}] and [\ref{f_fig2b}] where we plot the ratio of the dipole to higher multipoles at an age of $10^7$~years.
It is seen that the results are insensitive to these parameters, particularly for low orders of multipoles of interest.\\

\bef
\begin{center}{\mbox{\epsfig{file=figure2a.ps,width=225pt}}}\end{center}
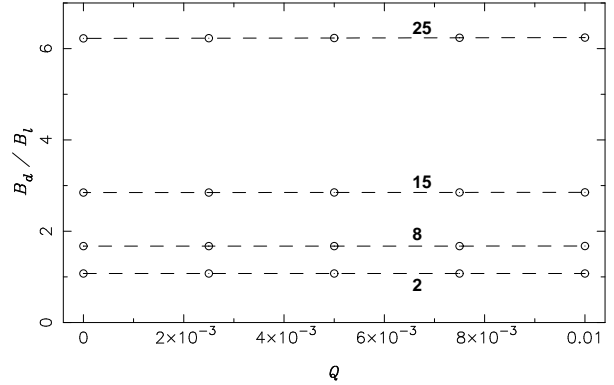
\caption[]{The ratio of the dipole surface field to that of the multipoles at $10^7$~years as a function of $Q$. The numbers next to the curves correspond to respective orders of multipole.
All curves correspond to a depth of $x = 0.98$ corresponding to a density of $\rho = 10^{11}~\gcc$, at which the initial current 
is concentrated.}
\label{f_fig2a}
\eef

\bef
\begin{center}{\mbox{\epsfig{file=figure2b.ps,width=225pt}}}\end{center}
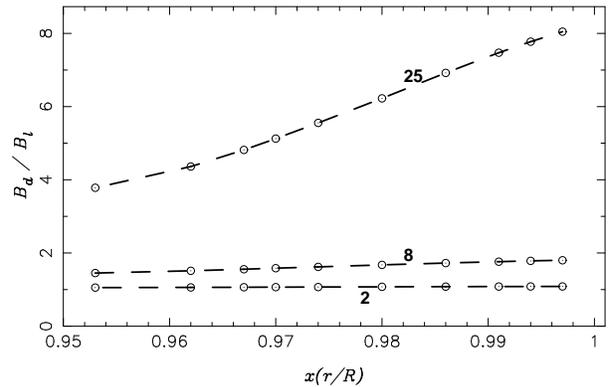
\caption[]{The ratio of the dipole surface field to that of the multipoles at $10^7$~years as a function of depth. The points 
marked in the plots here correspond to densities $\rho = 10^{13.5}, 10^{13}, 10^{12.5}, 10^{12}, 10^{11.5}, 10^{11}, 10^{10.5}, 
10^{10}, 10^{9.5}, 10^{9}~\gcc$. The numbers next to the curves correspond to respective 
orders of multipole. All curves correspond to $Q = 0$.}
\label{f_fig2b}
\eef

\ni Krolik (1991) and Arons (1993) conjectured that except for multipoles of order $l \gsim R_{\ast}/\triangle r$ the decay 
rates would be similar due to the finite thickness $\triangle r$ of the crust over which the current is confined. The 
evolution plotted in figure [\ref{f_fig1a}] assumes that $\triangle r = 1.2$~km for which $R_{\ast}/\triangle r \sim 8$. 
However it is seen from figures [\ref{f_fig1a}] and [\ref{f_fig1b}] that significant decay occurs only for $l\gsim 25$, much greater 
than $R_{\ast}/\triangle r$. This is most likely caused by steep increase in conductivity towards the interior.\\

\ni In conclusion, our results indicate that for a crustal model of the neutron star magnetic field there would be no significant 
change in the multipolar structure with age. This fact seems to be corroborated by observations: studies identifying multiple 
components in pulse profiles (Kramer \etal, 1994) show that the number of components does not vary with the age of the pulsar. Thus 
the evolution of the multipolar structure of the magnetic field is unlikely to leave any observable signature on pulsar emission.
This is in contrast with the predictions from the plate-tectonics model of Ruderman (1991a,b,c) which suggests a major change in 
the field structure with pulsar spin evolution.

\section*{Acknowledgment}
We thank A.~A. Deshpande, Rajaram Nityananda, N. Rathnasree, 
V. Radhakrishnan, C. Shukre and M. Vivekanand for helpful discussions. 
We are grateful to V. Urpin for providing us the computer-readable versions
of cooling curves computed by Urpin and van Riper (1993).
We gratefully acknowledge D. Page for bringing the X-ray work to our attention 
and an anonymous referee for his useful remarks.

\beb
\bi Arons J., 1993, ApJ, 408, 160
\bi Arons J., 1998, To be published in `Neutron Stars and Pulsars', S. Shibata and M. Sato,eds.(Tokyo: Universal Academy Press), astro-ph/9802198
\bi Barnard J.~J., Arons J., 1982, ApJ, 254, 713
\bi Baym G., Pethick C., Sutherland P., 1971, ApJ, 170, 299
\bi Bhattacharya D., Datta B., 1996, MNRAS, 282, 1059
\bi Bhattacharya D., Wijers R. A. M. J., Hartman J. W. \& Verbunt F. 1992, A\&A, 254, 198 
\bi Blandford R.~D., Applegate J.~H., Hernquist L. 1983, MNRAS, 204, 1025
\bi Deshpande A.~A., Rankin J.~M., 1998, BAAS, 193, 93.06
\bi Geppert U., Urpin V.~A., 1994, MNRAS, 271, 490
\bi Gudmundsson E.~H., Pethick C.~J., Epstein R.~J., 1983, ApJ, 272, 286
\bi Hartman J. W., Verbunt F., Bhattacharya D., Wijers R. A. M. J., 1997, A\&A, 322, 477
\bi Itoh N., Kohyama Y., Matsumoto N., Seki M., 1984, ApJ, 285, 758
\bi Jackson J.~D., 1975, {\em Classical Electrodynamics}, 2nd ed., John Wiley \& Sons
\bi Konar S., 1997, {\em Ph. D. thesis}, Indian Institute of Science, Bangalore
\bi Konar S., Bhattacharya D., 1997, MNRAS, 284, 311
\bi Konar S., Bhattacharya D., 1999, MNRAS, {\em in press} 
\bi Kramer M., Wielebinski R., Jessner A., Gil J.~A., Seiradakis J.~H., 1994, A\&AS, 107, 515
\bi Krolik J.~H., 1991, ApJ, 373, L69
\bi Mukherjee S., Kembhavi A., 1997, ApJ, 489, 928
\bi Negele J.~W., Vautherin D., 1973, Nucl. Phys. A, 207, 298
\bi Page D., Sarmiento A., 1996, ApJ, 473, 1067
\bi Rankin J.~M., Deshpande A.~A., 1998, BAAS, 193, 41.08
\bi Rankin J.~M., Rathnasree N., 1995, JAA, 16, 327
\bi Ruderman M., 1991a, ApJ, 366, 261
\bi Ruderman M., 1991b, ApJ, 382, 576
\bi Ruderman M., 1991c, ApJ, 382, 587
\bi Ruderman M., Sutherland P.~G., 1975, ApJ, 196, 51
\bi Sang Y., Chanmugam G., 1987, ApJ, 323, L61
\bi Seiradakis J.~H., 1998, {\em private communications}
\bi Srinivasan G., Bhattacharya D., Muslimov A.~G., Tsygan A.~I., 1990, Curr. Sc., 59, 31
\bi Urpin V.~A., Geppert U., 1995, MNRAS, 275, 1117
\bi Urpin V.~A., Geppert U., 1996, MNRAS, 278, 471
\bi Urpin V.~A., van Riper K.~A., 1993, ApJ, 411, L87
\bi van Riper K.~A., 1991, ApJS, 75, 449
\bi Vivekanand M., Radhakrishnan V., 1980, JAA, 1, 119
\bi Wendell C.~K., Van Horn H.~M., Sargent D., 1987, ApJ, 313, 284
\bi Wiringa R.~B., Fiks V., Fabrocini A., 1988, Phys. Rev. C, 38, 1010
\bi Yakovlev D.~G., Urpin V.~A., 1980, SvA, 24, 303
\eeb

\end{document}